\documentclass[sigconf]{acmart}
\acmConference[ESEC/FSE 2017]{Submitted to 11th Joint Meeting of the European Software Engineering Conference and the ACM SIGSOFT Symposium on the Foundations of Software Engineering}{4--8 September, 2017}{Paderborn, Germany}

\usepackage{booktabs} 
\usepackage{algorithm, algorithmicx} 
\usepackage{algpseudocode}
\usepackage{multirow}
\usepackage{tabu}
\usepackage{enumitem} 

\usepackage[framemethod=tikz]{mdframed}
\usepackage{lipsum}
\usetikzlibrary{shadows}
\usepackage{mathtools}
\usepackage{balance}
\usepackage{graphics}
\newmdenv[
tikzsetting= {fill=white},
linewidth=1pt,
roundcorner=2pt, 
shadow=false
]{myshadowbox}

\usepackage{color,colortbl}
\usepackage{xcolor}
\definecolor{lightgray}{gray}{0.8}

\makeatletter
\let\th@plain\relax
\makeatother

\usepackage[framed]{ntheorem}
\usepackage{framed}
\usepackage{tikz}
\usetikzlibrary{shadows}

\definecolor{Gray}{rgb}{0.88,1,1}
\definecolor{Gray}{gray}{0.85}
\theoremclass{Lesson}
\theoremstyle{break}
\tikzstyle{thmbox} = [rectangle, rounded corners, draw=black,
 fill=Gray,  drop shadow={fill=black, opacity=1}]

\newcommand{\bi}{\begin{itemize}[leftmargin=0.4cm]}
\newcommand{\ei}{\end{itemize}}
\newcommand{\be}{\begin{enumerate}}
\newcommand{\ee}{\end{enumerate}}

\newcommand{\fig}[1]{Figure~\ref{fig:#1}}
\newcommand{\tab}[1]{Table~\ref{tab:#1}}

\newshadedtheorem{lesson}{Result}

\setcopyright{rightsretained}

\acmConference[FSE'2017]{ACM SIGSOFT SYMPOSIUM ON THE FOUNDATIONS OF SOFTWARE ENGINEERING}{September 2017}{PADERBORN, GERMANY} 
\acmYear{2017}
\copyrightyear{2017}

\acmPrice{15.00}

\setcopyright{none}
\acmDOI{10.1145/3106237.3106256}
\acmISBN{978-1-4503-5105-8/17/09}
\acmConference[ESEC/FSE'17]{2017 11th Joint Meeting of the European Software Engineering Conference and the ACM SIGSOFT Symposium on the Foundations of Software Engineering}{September 4-8, 2017}{Paderborn, Germany} 
\acmYear{2017}
\copyrightyear{2017}
\acmPrice{15.00}

\begin{document} 
\title{Easy over Hard: A Case Study on Deep Learning}

\author{Wei Fu,  Tim Menzies}
\affiliation{%
  \institution{ Com.Sci., NC State, USA}
}
\email{wfu@ncsu.edu, tim.menzies@gmail.com}

 
\begin{abstract}
While deep learning is an exciting new technique, the benefits of this
method need to be  assessed with respect to its computational cost. 
This is particularly important for deep learning since these learners need 
 hours (to  weeks) to train the model.
Such long training time limits the ability
of (a)~a researcher to test
the stability of their conclusion
via repeated runs with different random seeds;
and (b)~other researchers to repeat, improve, or even refute that original work.

For example, recently, deep learning
was used to  find which
questions  in the Stack Overflow programmer discussion
forum can be linked together. That deep learning system
took 14 hours to execute.
We show here that applying a very simple optimizer called DE to fine tune SVM, it can achieve  similar (and sometimes better) results. The DE approach  terminated in 10 minutes;
i.e. 84 times faster 
hours than  deep learning method.

We offer these results as a cautionary tale to the software analytics community and suggest that not
every new innovation should be applied without critical analysis. If researchers deploy some new and expensive process, that work should be baselined against some simpler and faster alternatives.

\end{abstract}

%
%




\maketitle


\section{Introduction}

This paper extends a prior result from ASE'16 by
 Xu et al.~\cite{xu2016predicting} (hereafter, XU). XU described
 a method to
 explore large programmer discussion forums, then 
 uncover related, but separate, entries.
 This is an important problem. Modern SE
 is evolving so fast that these  forums
 contain more relevant and recent comments on 
 current technologies than any textbook or research article.
 
 In their work, XU predicted whether two questions posted on Stack Overflow are semantically linkable. 
Specifically, XU define a question along with its entire set of answers posted on Stack Overflow
as a {\it knowledge unit}~(KU). If two knowledge units are semantically related, they are considered
as {\it linkable} knowledge units. 

In their paper, they used a convolution neural
network~(CNN), a kind of deep learning method~\cite{lecun2015deep}, to predict whether two KUs are linkable. Such 
CNNs are highly computationally expensive,
often requiring network composed of 10 to 20 layers, hundreds of millions of weights and billions of connections between units~\cite{lecun2015deep}. Even with
advanced hardware and algorithm parallelization, training deep learning models still requires hours to weeks.
For example:
\bi
\item
XU report that their analysis
required 14 hours of CPU.
\item
Le~\cite{le2013building} used  a cluster with 1,000 machines (16,000 cores) for three days to train a deep learner.
\ei
 
This paper debates what methods should be recommended
to those wishing to repeat the analysis of XU. We focus on whether using simple and faster methods can achieve
the results that are currently achievable by the state-of-art deep learning method.
Specifically, we repeat XU's
study using  DE  (differential evolution~\cite{storn1997differential}),
which serves as a hyper-parameter optimizer to tune XU's baseline method, which is a conventional machine learning algorithm, support vector machine~(SVM).
Our study asks:

{\it \textbf{RQ1}: Can we reproduce XU's baseline results (Word Embedding + SVM)?}
Using such a baseline, we can compare our methods to those of XU.
 
 {\it \textbf{RQ2}: Can   DE   tune a standard learner such that
 it outperforms
  XU's deep learning method?}
 We apply differential evolution  to tune SVM. In terms of precision, recall and F1-score, we observe that the tuned SVM method outperforms CNN in most evaluation scores.

{\it \textbf{RQ3}: Is tuning SVM with DE faster than XU's deep learning method?}
Our   DE method
 is  $84$ times faster than CNN. 
 
We offer these results as a cautionary tale to the software analytics community.
While deep learning is an exciting new technique, the benefits of this
method need to be carefully assessed with respect to its computational cost. 
More generally,
if researchers deploy some new and expensive process (like deep learning), that work should be baselined against some simpler and faster alternatives

The rest of this paper is organized as follows. Section~\ref{background} describes the background and related work on deep learning and parameter tuning in SE. Section~\ref{method} explains the case study problem and the proposed tuning method investigated in this study, then Section~\ref{experiment} describes the
experimental settings of our study, including research questions, data sets, evaluation measures and experimental design.
Section~\ref{results} presents the results. Section~\ref{discussion} discusses implications from the results and the threats to the validity of our study. Section~\ref{conclusion} concludes the paper and discusses the future work.

Before beginning, we digress to make two points.
Firstly, just because ``DE + SVM'' beats deep learning
in this application, this does not mean DE is always the superior method for all other software analytics applications. No learner works best over all problems~\cite{wolpert1996lack}-- the trick is to try several approaches and select the one that works best on the local data. Given the low computational cost of DE (10 minutes vs 14 hours), DEs are an obvious and low-cost candidate for exploring such alternatives. 

Secondly, to enable other researchers to repeat, improve, or
refute our results, all our scripts and data are 
freely available on-line 
Github\footnote{\url{https://github.com/WeiFoo/EasyOverHard}}.

\section{Background and Related Work}\label{background}

\subsection{Why Explore Faster Software Analytics?}
This section argues that
avoiding slow methods for software analytics is an 
open and urgent issue.

Researchers and industrial practitioners now routinely make extensive use of software analytics to discover (e.g.) how long it will take to integrate the new code~\cite{czerwonka2011crane}, where bugs are most likely to occur~\cite{ostrand2004bugs}, who should fix the bug~\cite{anvik2006should}, or how long it will take to develop their code~\cite{kocaguneli2012value,kocaguneli2012exploiting,molokken2003review}.  Large organizations like Microsoft routinely practice data-driven policy development where organizational policies are learned from an extensive analysis of large data sets collected from developers~\cite{begel2014analyze,theisen2015approximating}.

But the more complex the  method,
the harder it is to apply the analysis.
Fisher et al.~\cite{fisher2012interactions} characterizes software analytics as a work flow that distills large quantities of low-value data down to smaller sets of higher value data. Due to the complexities and computational cost of SE analytics, ``the luxuries of interactivity, direct manipulation, and fast system response are gone''~\cite{fisher2012interactions}. They characterize modern cloud-based analytics as a throwback to the 1960s-batch processing mainframes where jobs are submitted and then analysts wait, wait, and wait for results with ``little insight into what is really going on behind the scenes, how long it will take, or how much it is going to cost''~\cite{fisher2012interactions}. Fisher et al. ~\cite{fisher2012interactions} document the issues seen by 16 industrial data scientists, one of whom remarks \begin{quote}
``Fast iteration is key, but incompatible with the jobs are submitted and processed in the cloud. It is frustrating to wait for hours, only to realize you need a slight tweak to your feature set''.
\end{quote}

Methods for improving the quality of modern software analytics have made this issue even more serious. There has been continuous development of new feature selection~\cite{hall2003benchmarking} and feature discovering~\cite{jiang2013personalized} techniques for software analytics, with the most recent ones focused on deep learning methods. These are all exciting innovations with the potential to dramatically improve the quality of our software analytics tools. Yet these are all CPU/GPU-intensive methods. For instance:
\bi
\item Learning control settings for learners can take days to weeks to years of CPU time~\cite{fu2016differential,tantithamthavorn2016automated,wang2013searching}.
\item Lam et al. needed weeks of CPU time to combine deep learning and text mining to localize buggy
files from bug reports~\cite{lam2015combining}.
\item Gu et al. spent $240$ hours of GPU time  to train a deep learning based method to generate API usage sequences for given natural language query~\cite{gu2016deep}. 
\ei 
Note that the above problem is not solvable by waiting for faster CPUs/GPUs. We 
can no longer rely on Moore's Law ~\cite{moore1998cramming} to double our computational power every 18 months. Power consumption and heat dissipation issues effect block further exponential increases to CPU clock frequencies~\cite{kumar2003single}. Cloud computing environments are extensively monetized so the total financial cost of training models can be prohibitive, particularly for long running tasks. For example, it would take 15 years of CPU time to learn the tuning parameters of software clone detectors proposed in ~\cite{wang2013searching}. Much of that CPU time can be saved if there is a faster  way.

\subsection{What is Deep Learning?}

Deep learning is a branch of machine learning built on multiple layers of neural networks that attempt to model high level abstractions in data. According to LeCun et al.~\cite{lecun2015deep}, deep learning methods are representation-learning methods with multiple levels of representation,
obtained by composing simple but non-linear modules that each
transforms the representation at one level (starting with the raw input)
into a representation at a higher, slightly more abstract level. Compared to the conventional
machine learning algorithms, deep learning methods are very good at exploring high-dimensional data.

By utilizing extensive computational power, 
deep learning has been proven to be a very powerful method 
by researchers in many fields~\cite{lecun2015deep}, like computer vision and natural language processing~\cite{krizhevsky2012imagenet,mikolov2013distributed,sutskever2014sequence,schmidhuber2015deep,arel2010deep}. In 2012, 
Convolution neural networks method won the ImageNet competition ~\cite{krizhevsky2012imagenet},
which achieves half of the error rates of the best competing
approaches. After that, CNN became the dominant approach for almost all recognition
and detection tasks in computer vision community. CNNs are designed to process the data in the form of multiple arrays, e.g., image data. According to LeCun et al.~\cite{lecun2015deep},
recent CNN methods are usually a huge network composed of 10 to 20 layers, hundreds of millions of weights and billions of connections between units. With advanced hardware and algorithm parallelization, training such model still need a few hours~\cite{lecun2015deep}. 
For the tasks that deal with sequential data, like text and speech, recurrent neural networks~(RNNs) have been shown to work well. 
RNNs are found to be good at predicting the next character or word given the context. For example, Graves et al.~\cite{graves2013speech} proposed to use long short-term memory~(LSTM) RNNs to perform speech recognition, 
which achieves a test set error of $17.7\%$ on the benchmark testing data. Sutskever et al.~\cite{sutskever2014sequence} used two multiplelayered LSTM RNNs to translate sentences in English to French.

\subsection{Deep Learning in SE}

We study deep learning since, recently, it  has attracted 
much attentions from researchers and practitioners in software
 community~\cite{wang2016automatically, gu2016deep, xu2016predicting,white2016deep,white2015toward,lam2015combining,choetkiertikul2016deep,yuan2014droid,mou2016convolutional}.
 These researchers applied  deep learning techniques to solve various problems,
 including defect prediction, bug localization, clone code detection, malware detection, API recommendation, 
 effort estimation and linkable knowledge prediction.
 
We find that this work   can be divided into   two categories:
 
\bi
\item Treat deep learning as a feature extractor, and then apply other  machine learning algorithms to do further work~\cite{lam2015combining,wang2016automatically,choetkiertikul2016deep}.
\item Solve problems directly with  deep learning~\cite{gu2016deep,xu2016predicting,white2016deep,white2015toward,yuan2014droid,mou2016convolutional}.
\ei
\noindent
\subsubsection{Deep Learning as  Pre-Processor}

Lam et al.~\cite{lam2015combining}  proposed an approach to apply deep neural network
 in combination with rVSM to automatically locate the potential buggy files for a given
 bug report. By comparing it to baseline methods~(Naive Bayes~\cite{kim2013should}, learn-to-rank~\cite{ye2014learning}, 
 BugLocator~\cite{zhou2012should}), Lam et al. reported, $\mbox{16.2-46.4}\%$, $\mbox{8-20.8}\%$  and $\mbox{2.7-20.7}\%$ 
 higher top-1 accuracy than baseline methods, respectively~\cite{lam2015combining}. The training time for deep neural
 network was reported from 70 to 122 minutes for 6 projects on a computer with 32 cores 2.00GHz CPU,
 126 GB memory. However,
 the runtime information of the baseline methods was not reported.
 
 Wang et al.~\cite{wang2016automatically} applied deep belief network to automatically
 learn semantic features from token vectors extracted from the studied software program. After
 applying deep belief network to generate features from software code,
 Naive Bayes, ADTree and Logistic Regression methods are used to evaluate the effectiveness
 of feature generation, which is compared to the same learners using traditional 
 static code features~(e.g.  McCabe metrics, Halstead's effort metrics and  CK object-oriented code mertics~\cite{kafura1987use,chidamber1994metrics,mccabe1976complexity,halstead1977elements}). In terms of
 runtime, Wang et al. only report time for generating semantics features with deep belief network, which
 ranged from 8 seconds to 32 seconds~\cite{wang2016automatically}. However, the time for training and tuning deep belief network is
 missing. Furthermore, to compare the effectiveness of deep belief network for generating features with methods that extract traditional 
 static code features in terms of time cost, 
 it would be favorable to include all the time spent on feature extraction, including
 paring source code, token generation and token mapping for both deep belief network and traditional methods~(i.e., an end-to-end comparison).
 
 Choetkiertikul et al.~\cite{choetkiertikul2016deep} proposed to apply deep learning techniques
 to solve effort estimation problems on user story level. 
 Specifically,  Choetkiertikul et al. ~\cite{choetkiertikul2016deep} proposed to leverage
 long short-term memory~(LSTM) to learn feature vectors from the title,
 description and comments associated with an issue report and after that,
 regular machine learning techniques, like CART, Random Forests, 
 Linear Regression and Case-Based Reasoning are applied to build the effort
 estimation models. Experimental results show that
 LSTM has a 
 significant improvement over the baseline method bag-of-words.
 However, no further information regarding
 runtime as well as experimental hardware  is reported for both methods and there is no cost 
 of this deep learning method at all.

\noindent
\subsubsection{Deep Learning as  a Problem Solver}

 White et al.~\cite{white2015toward, white2016deep} applied
 recurrent neural networks, a type of  deep learning techniques, 
 to address code clone detection and code suggestion. They reported,
 the average training time for 8 projects were ranging from 34 seconds
  to 2977 seconds for each epoch on a computer with two 3.3 GHz
 CPUs and each project required at least 30 epochs~\cite{white2016deep}.
Specifically, for the {\it JDK} project in their experiment, it would take 25 hours 
 on the same computer to train the models before getting prediction.
 For the time cost for code suggestions, authors did not mention any related information~\cite{white2015toward}.

Gu et al.~\cite{gu2016deep} proposed  a recurrent neural network~(RNN)
 based method, D{\scriptsize EEP}API, to generate API usage sequences for a given natural language query. 
 Compared with the baseline method { SWIM}~\cite{raghothaman2016swim} and 
 { Lucene + UP-Miner}~\cite{wang2013mining},  D{\scriptsize EEP}API   improved the performance significantly.
 However, that improvement came at a cost: that  model was trained with a Nivdia K20 GPU for  240 hours~\cite{gu2016deep}.
 
 XU~\cite{xu2016predicting} utilized neural language model and  
 convolution neural network~(CNN) to  learn word-level and document-level features to
 predict semantically linkable knowledge units on Stack Overflow. 
 In terms of performance metrics, like precision, recall and F1-score,
 CNN method was evaluated much better than 
 the baseline method support vector machine~(SVM). 
 However, once again, that performance improvement came at a cost:
 their deep learner required  
 14 hours to train CNN model on a 2.5GHz PC with 16 GB RAM~\cite{xu2016predicting}.
 
 Yuan et al.~\cite{yuan2014droid} proposed a deep belief network based method for
 malware detection on Android apps. By training and testing
 the deep learning model with  200 features extracted
 from static analysis and dynamic analysis from 500 sampled Android app, they
 got $96.5\%$ accuracy for deep learning method  and $80\%$ for one baseline method, SVM~\cite{yuan2014droid}.
 However, they did not report any runtime comparison between the deep learning method and 
 other classic machine learning methods.
 
 Mou et al.~\cite{mou2016convolutional} proposed a  tree-based convolutional neural network
 for programming language processing, in
which a convolution kernel is designed over programs' abstract
syntax trees to capture structural information. Results show that their method achieved
$94\%$ accuracy, which is better than the baseline method RBF SVM $88.2\%$
on program classification problem~\cite{mou2016convolutional}. However, Mou
et al.~\cite{mou2016convolutional} did not discuss any runtime comparison between the proposed method and
baseline methods.

 \subsubsection{Issues with Deep Learning}
 In summary,  deep learning is used extensively in software
 engineering community.  A common pattern
 in that research is to:
 \bi
 \item
 Report deep learning's benefits, but not  its CPU/GPU cost~\cite{white2015toward,choetkiertikul2016deep,yuan2014droid,mou2016convolutional};
 \item
 Or simply show the cost, without further analysis~\cite{wang2016automatically, lam2015combining, gu2016deep, xu2016predicting, white2016deep}.
 \ei
Since  deep learning techniques cost large amount of time and computational
resources to train its model,
one might question whether the improvements from deep learning is worth
the costs. {\it Are there any simple techniques that achieve similar improvements
with less resource costs?} To investigate how simple methods could improve baseline
methods, we select XU~\cite{xu2016predicting} study as a case study. The reasons
are as follows:
\bi 
 
\item Most deep learning paper's baseline methods in SE are either not publicly available or too complex to implement~\cite{white2016deep,lam2015combining}. XU define their baseline methods precisely enough so others can confidently reproduce it locally.
XU's baseline method is SVM learner, which is available in many machine learning toolboxes.

\item
Further, it is not
yet common practice for deep learning researchers in SE community
to share their implementations and data~\cite{white2016deep,white2015toward,lam2015combining,wang2016automatically,choetkiertikul2016deep,gu2016deep}, where a tiny difference may lead to a huge difference in the results. Even though XU do not share their CNN tool, their
training and testing data are available online, which can be used for our proposed method. Since the same training and testing data are used for XU's CNN and our proposed method,  we can compare results of our method to their CNN results.
\item Some studies do not report their runtime and experimental environment, which makes it harder for us
to systematically compare our results with theirs in terms of computational costs~\cite{choetkiertikul2016deep,yuan2014droid,white2015toward, mou2016convolutional}.
XU clearly report their experimental hardware and runtime, which will be easier for us compare our computational costs to theirs.
\ei


\subsection{Parameter Tuning in SE}\label{tuning}
In this paper, we use DE as an optimizer to do parameter tuning for SVM, which
achieves results that are competitive with deep learning. This section
discusses related work on parameter tuning in SE community.

Machine learning algorithms are designed to explore the instances
to learn the bias. However, most of these algorithms are controlled by parameters
such as:

\bi
\item
The maximum allowed depth
of decision tree built by CART;
\item
The number of trees to be built within a Random Forest.
\ei

Adjusting these parameters
is called hyperparameter optimziation.
It is a well   well explored approach in other communities ~\cite{bergstra2012random,li2016hyperband}. However, in SE,
such parameter optimization is not a common task~(as shown in the following examples).

In the field of {\em defect prediction}, 
Fu et al.~\cite{fu2016tuning}  surveyed hundreds of highly 
cited software engineering paper about defect prediction. 
Their observation is that most software engineering  researchers
did not acknowledge the impact of tunings 
~(exceptions:~\cite{lessmann2008benchmarking,tantithamthavorn2016automated}) and
use the ``off-the-shelf'' data miners. For example, 
Elish et al.~\cite{elish2008predicting} compared support vector machines
to other data miners for the purposes of defect prediction.
However, the Elish et al. paper makes no mention of any SVM tuning study~\cite{elish2008predicting}. More details about their survey 
refer to~\cite{fu2016tuning}.

In the field of {\em topic modeling}, Agrawal et al.~\cite{agrawal2016wrong} investigated 
the impact of parameter tuning on Latent Dirichlet Allocation~(LDA).
LDA 
  is a widely used technique in software engineering field
to find related topics within unstructured text, 
like topic analytics on Stack Overflow ~\cite{barua2014developers}
and source code analysis ~\cite{binkley2014understanding}.
 Agrawal et al. found that LDA suffers from conclusion instability
~(different input orderings can lead to very different results)
that is a result of poor choice of the LDA control parameters~\cite{agrawal2016wrong}. 
Yet, in their survey of LDA use in SE, they found that very few  
researchers ~(4 out of 57 papers) explored the benefits of parameter tuning for LDA.

One troubling trend is that, in the few SE papers that perform tuning,
they do so using methods heavily deprecated in the machine learning community.
For example, two SE papers that use tuning~\cite{lessmann2008benchmarking,tantithamthavorn2016automated}, 
apply a simple grid search to explore the potential parameter space for optimal tunings
~(such grid searchers run one for-loop for each parameter being optimized). 
However, Bergstra et al.~\cite{bergstra2012random} and 
Fu et al.~\cite{fu2016differential} argue that random search methods
~(e.g. the 
differential evolution algorithm used here) are better than 
grid search in terms of efficiency and performance.


\section{Method}\label{method}

\begin{table}[!htp]
\caption{Classes of Knowledge Unit Pairs.}\label{tab:classes}
\centering
\begin{tabular}{l|l}
\hline
\rowcolor{lightgray}
Class & Description  \\\hline
\multirow{2}{*}{Duplicate} & These two knowledge units are addressing the\\
                          & same  question.\\ \hline
\multirow{2}{*}{Direct link}& One knowledge unit can help to  answer the\\ 
                         &  question in the other  knowledge unit.\\ \hline
\multirow{3}{*}{Indirect link}& One knowledge provides similar information to\\
                         &  solve the  question in the other knowledge unit, \\
                         &  but not a direct answer.\\ \hline
\multirow{2}{*}{Isolated}& These two knowledge units discuss unrelated \\
                         & questions.\\ \hline
\end{tabular}
\end{table}

\subsection{Research Problem}\label{problem}

This section is an overview of the the task and methods 
used by XU. Their task was to  predict relationships between two knowledge units (questions with answers) on Stack Overflow. Specifically, 
XU  divided linkable knowledge  unit pairs
into 4 difference categories namely, {\it duplicate}, {\it direct link}, {\it indirect link} and {\it isolated}, 
based on its relatedness. The definition of these four categories are shown in \tab{classes}~\cite{xu2016predicting}:


In that paper, XU provided the following two methods as baselines~\cite{xu2016predicting}:

\bi
\item TF-IDF + SVM: a multi-class SVM classifier with  36 textual features generated  based on the 
TF and IDF values of the words in a pair of knowledge units. 
\item Word Embedding + SVM:  a multi-class SVM classifier with word embedding generated by the word2vec model~\cite{mikolov2013distributed}.
\ei
Both of these two baseline methods are compared against their proposed method, Word Embedding + CNN. 

In this study, we select  Word Embedding + SVM as the baseline because it uses word embedding as the input,
which is the same as the Word Embedding + CNN method by XU.

\begin{table*}[htp]
   \caption {List of Parameters Tuned by This Paper.}
\centering
\resizebox{\textwidth}{!}{
	\begin{tabular}{|c|c|c|c|l|}
	\cline{1-5}
	Parameters & Default &Xue et al.&Tuning Range& 
\multicolumn{1}{c|}{Description} \\ \hline
	C & 1.0 &unknown&[1, 50]& Penalty parameter C of the error term.\\ \cline{1-5} 
	 kernel & `rbf' &`rbf'&[`liner', `poly', `rbf', `sigmoid']& Specify the kernel type to be used in the algorithms. \\ \cline{1-5} 
	 gamma & {1/n\_features} &$1/200$& [0, 1]& Kernel coefficient for `rbf', `poly' and `sigmoid'. \\ \cline{1-5} 
	 coef0 & 0 & unknown & [0, 1] &  Independent term in kernel function. It is only used in `poly' and `sigmoid'. \\ \cline{1-5}
\hline
	\end{tabular}}
\label{tab:parameters}
\end{table*}

\subsection{Learners and Their Parameters}
SVM has been proven to be a very successful method to solve
text classification problem. A SVM  seeks to minimize misclassification
errors by selecting a boundary or hyperplane that leaves
the maximum margin between positive and negative classes~(where the
margin is defined as the sum of the distances of the
hyperplane from the closest point of the two classes~\cite{joachims1998text}).

Like most machine learning algorithms, there are some parameters associated with
SVM to control how it learns.  In XU's experiment, they used a radial-bias function~(RBF) for their SVM kernel
and set $\gamma$ to $1/k$, where $k$ is $36$ for TF-IDF + SVM method
and $200$ for Word Embedding + SVM method. For other parameters, 
XU mentioned that grid search was applied to optimize the SVM parameters, 
but no further information was disclosed. 

For our work, we used the SVM module from Scikit-learn~\cite{scikit-learn}, a Python package for machine learning,
where the parameters shown in Table. ~\ref{tab:parameters} are selected for tuning.
Parameter $\mathit{C}$ is to set the amount of regularization, which controls the tradeoff between
the errors on training data and the model complexity.  A small value for {\it C} will generate 
a simple model with more training errors, while a large value will lead to a complicated model with fewer
errors. {\it Kernel} is to introduce different nonlinearities into the SVM model by applying kernel functions
on the input data. {\it Gamma } defines how far the influence of a single training example reaches, 
with low values meaning `far' and high values meaning `close'. {\it coef0} is an independent parameter used
in sigmod and  polynomial kernel function.

As to why we used the ``Tuning Range'' shown in \tab{parameters}, and not some other ranges,
we note that (a)~those ranges include the defaults and also XU's values; (b)~the results presented below
show that by exploring those ranges,  we achieved large gains in the performance of our baseline method.
This is not to say that {\em larger} tuning ranges might not result in {\em greater} improvements.
However, for the goals of this paper (to show that tuning baseline method does matter), exploring
just these ranges shown in \tab{parameters} will suffice.


\subsection{Learning Word Embedding}\label{embedding}
Learning word embeddings refers to find vector representations
of words such that the similarities between words can be captured by cosine similarity of corresponding 
vector representations. It is been shown that the words with similar semantic and syntactic are found closed
to each other in the embedding space~\cite{mikolov2013distributed}.

Several methods have been proposed to generate word embeddings, 
like skip-gram~\cite{mikolov2013distributed}, GloVe ~\cite{pennington2014glove}
and PCA on the word co-occurrence matrix~\cite{lebret2013word}. To replicate XU work,
we used the continuous skip-gram model~(word2vec),  which is a unsupervised word representation learning method based on
neural networks and also used by  XU~\cite{xu2016predicting}. 

The skip-gram model learns vector representations of words
 by predicting the surrounding words in a context window. 
 Given a sentence of words $W =w_1$,$w_2$,...,$w_n$, the objective of skip-gram model is to maximize the
 the average log probability of the surrounding words:
 \begin{equation*}
 \frac{1}{n}\sum_{i=1}^{n} \sum_{-c\leq j \leq c, j \neq 0} log p(w_{i+j}|w_i)
\end{equation*}
where $c$ is the context window size and $w_{i+j}$ and $w_{i}$ represent surrounding words and center word, respectively.
The probability of $p(w_{i+j}|w_i)$ is computed according to the softmax function:

\begin{equation*}
p(w_O|w_I) = \frac{exp(v_{w_O}^Tv_{w_I})}{\sum_{w=1}^{|W|}exp(v_{w}^Tv_{w_I})}
\end{equation*}
where $v_{w_I}$ and $v_{w_O}$ are the vector representations of the input and output vectors of $w$, respectively. 
$\sum_{w=1}^{|W|}exp(v_{w}^Tv_{w_I})$  normalizes the inner product results across all the words.
To improve the computation efficiency, Mikolove et al. ~\cite{mikolov2013distributed} proposed
hierachical softmax and negative sampling
techniques. More details can be found in Mikolove et al.'s study~\cite{mikolov2013distributed}.

\begin{figure}[!htp]\small
     \begin{tabular}{|p{.95\linewidth}|}\hline
      1. Given a model~(e.g., SVM) with $n$ decisions~(e.g., $n=4$),
      TUNER calls SAMPLE $N=10*n$ times.
      Each call generates one member of the population {\em pop$_{i\in N}$}.

      2. TUNER scores each {\em pop}$_i$ according to various objective
      scores $o$. In the case of our tuning SVM, the objective $o$ is to maximize
     {F1-score}

     3. TUNER tries to each replace {\em pop}$_i$ with a mutant $m$
     built using Storn's differential evolution method~\cite{storn1997differential}.
     DE extrapolates between three other members of population $a,b,c$.
     At probability $p_1$, for each decision $a_k \in a$, then
     $m_k= a_k \vee ~(p_1 < \mathit{rand}() \wedge( b_k \vee c_k))$.

     4. Each mutant $m$ is assessed by calling  $\text{EVALUATE}(\textit{model, prior=m})$;
     i.e. by seeing what can be achieved within a goal after first assuming
     that $\textit{prior}=m$.

     5. To test if the mutant $m$ is preferred to {\em pop}$_i$, TUNER simply
      compare SCORE($m$) with SCORE({\em pop}$_i$). In case of our tuning SVM,
      the one with higher score will be kept.

    6. TUNER repeatedly loops over the population, trying to replace  items with mutants, until new better mutants stop being found.

    7. Return the best one in the population as the optimal tunings.
    \\\hline
    \end{tabular}
    \caption{Procedure TUNER: strives to find ``good'' tunings which maximizes
     the objective score of the model on training and tuning data. TUNER is based on Storn's differential evolution optimizer~\cite{storn1997differential}.}
    \label{fig:optimize}
\end{figure}

Skip-gram's parameters control how that algorithm
  learns   word embeddings. Those parameters include
  {\it window size} and {\it dimensionality of embedding space}, etc. 
Zucoon et al.~\cite{zuccon2015integrating} found that embedding dimensionality
and context window size have no consistent impact on retrieval model performance. However,
Yang et al.~\cite{yang2016using} showed that large context window and dimension
 sizes are preferable to improve the performance when using CNN to solve  classification tasks
 for Twitter. Since this work is to compare performance of  tuning SVM  with CNN, where
 skip-gram model is used to generate word vector representations for both of these methods, 
 tuning parameter of skip-gram model is beyond the scope of this paper 
 (but we will explore it in future work).

To train our word2vec model, $100,000$ knowledge units tagged with ``java'' from
Stack Overflow {\it posts} table  (include titles, questions and answers)
are randomly selected as a word corpus\footnote{Without further explanation, 
all the experiment settings, including learner algorithms,
training/testing data split, etc, strictly follow XU's work. }. 
After applying proper data processing techniques proposed by XU, like
 remove the unnecessary HTML tags and keep short code snippets in
{\it code} tag, then fit the corpus into {\it gensim} word2vec module ~\cite{rehurek2010software},
which is a python wrapper over original word2vec package.

When converting knowledge units into vector representations, 
for each word $w_i$ in the post processed knowledge unit~(including title, question and answers),
we query the trained word2vec model to get the corresponding word vector representation $v_i$.
Then the whole knowledge unit with $s$ words
is converted to vector representation by element-wise addition, $Uv = v_i \oplus v_2 \oplus...\oplus v_s $. 
This vector representation is used
as the input data to SVM.

\subsection{Tuning Algorithm}

A tuning algorithm is an optimizer that  drives the learner to explore
the optimal parameter in a given searching space. According to our
literature review, there are several searching algorithms used in 
SE community:{\em 
simulated annealing}~\cite{feather2002converging,menzies2007data};
 various {\em genetic algorithms}~\cite{jones1996automatic,harman2007current, arcuri2011parameter} augmented by
techniques such as {\em differential evolution}
~\cite{storn1997differential, fu2016tuning, fu2016differential,chaves2015differential,agrawal2016wrong}, 
{\em tabu search} and {\em scatter search}~\cite{beausoleil2006moss,molina2007sspmo,corazza2013using};
{\em particle swarm optimization}~\cite{windisch2007applying}; 
numerous {\em decomposition} approaches that use
    heuristics to decompose the total space into   small problems,   then apply a
    {\em response surface methods}~\cite{krall2015gale};
     {\em NSGA-II} ~\cite{zhang2007multi}and {\em NSGA-III}~\cite{mkaouer2014high}.

Of all the mentioned algorithms,  the simplest are simulated annealing~(SA)  and 
differential evolution~(DE), each of which can be coded in less than a page of some high-level scripting language.
 Our reading of the current literature is that there are more  advocates for
differential evolution than SA. For example,  Vesterstrom and Thomsen~\cite{Vesterstrom04} found DE to be competitive with 
 particle swarm optimization and other GAs.  DEs have already been applied before for 
 parameter tuning in SE community to do parameter tuning~(e.g. see~\cite{omran2005differential, chiha2012tuning, fu2016tuning, fu2016differential, agrawal2016wrong}) .
Therefore, in this work, we adopt DE as our tuning algorithm and 
the main steps in DE is described in \fig{optimize}.

\section{Experimental Setup}\label{experiment}
\subsection{Research Questions}\label{RQ}
 To systematically investigate whether tuning can improve the 
 performance of baseline methods compared with deep learning method, we set
 the following three research questions:

 \bi
 \item {\it \textbf{RQ1}: Can we reproduce XU's baseline results (Word Embedding + SVM)?}
 \item {\it \textbf{RQ2}: Can   DE   tune a standard learner such that
 it outperforms XU's deep learning method?}
 \item {\it \textbf{RQ3}: Is tuning SVM with DE faster than XU's deep learning method?}
 \ei
 
 RQ1 is to investigate whether our implementation of Word Embedding + SVM method has
 the similar performance with XU's baseline, which makes sure that our following 
 analysis can be generalized to XU's conclusion. RQ2 and RQ3 lead us to
 investigate whether tuning SVM comparable with XU's deep learning from both 
 performance and cost aspects.

\subsection{Dataset and Experimental Design}
Our experimental data comes from Stack Overflow data dump of 
September 2016\footnote{https://archive.org/details/stackexchange},
where the {\it posts} table includes all the questions and answers posted on Stack Overflow
up to date and the {\it postlinks} table describes the relationships between posts, 
e.g., {\it duplicate} and {\it linked}. As mentioned in Section
\ref{problem}, we have four different types of relationships in knowledge unit pairs.
Therefore,  {\it linked} type is further divided into {\it indirectly linked} and {\it directly linked}.
Overall, four different types of data are generated according the following rules~\cite{xu2016predicting}:
\bi
\item Randomly select a pair of posts from the {\it postlinks} table, if the value
in  {\it PostLinkTypeId} field for this pair of posts is $3$, then this pair of posts is {\it duplicate} posts. 
Otherwise they're {\it directly linked} posts.

\item Randomly select a pair of posts from the {\it posts} table, if this pair of posts is linkable from each other according to
{\it postlinks} table and the distance between them are greater than 2~(which means they are not duplicate or directly linked posts), then this pair of posts is indirectly linked. If they're
not linkable, then this pair of posts is {isolated}.
\ei

In this work, we use the same training and testing
knowledge unit pairs as XU ~\cite{xu2016predicting}\footnote{\url{https://github.com/XBWer/ASEDataset}}, 
where 6,400 pairs of  knowledge units for training and 1,600 pairs for testing. And each type 
of linked knowledge units accounts for $1/4$ in both training and testing data. The reasons that
we used the same training and testing data as XU are:
\bi
\item It is to ensure that  performance of our baseline method is as closed to XU's as possible.
\item Since deep learning method is way complicated compared to SVM and a little difference in implementations
might lead to different results. To fairly compare with XU's result, we can use the  performance scores
of CNN method from XU's study~\cite{xu2016predicting} without any implementation bias introduced.
\ei

For training word2vec model, we randomly select 100,000 knowledge
 units~(title, question body and all the answers) from {\it posts} table that are
 related to ``java''. After that, all the training/tuning/testing knowledge units
 used in this paper are converted into word embedding representations by looking up
 each word in wrod2vec model as described in Section~\ref{embedding}.

 \begin{figure}
    \centering
     \includegraphics[width=.47\textwidth]{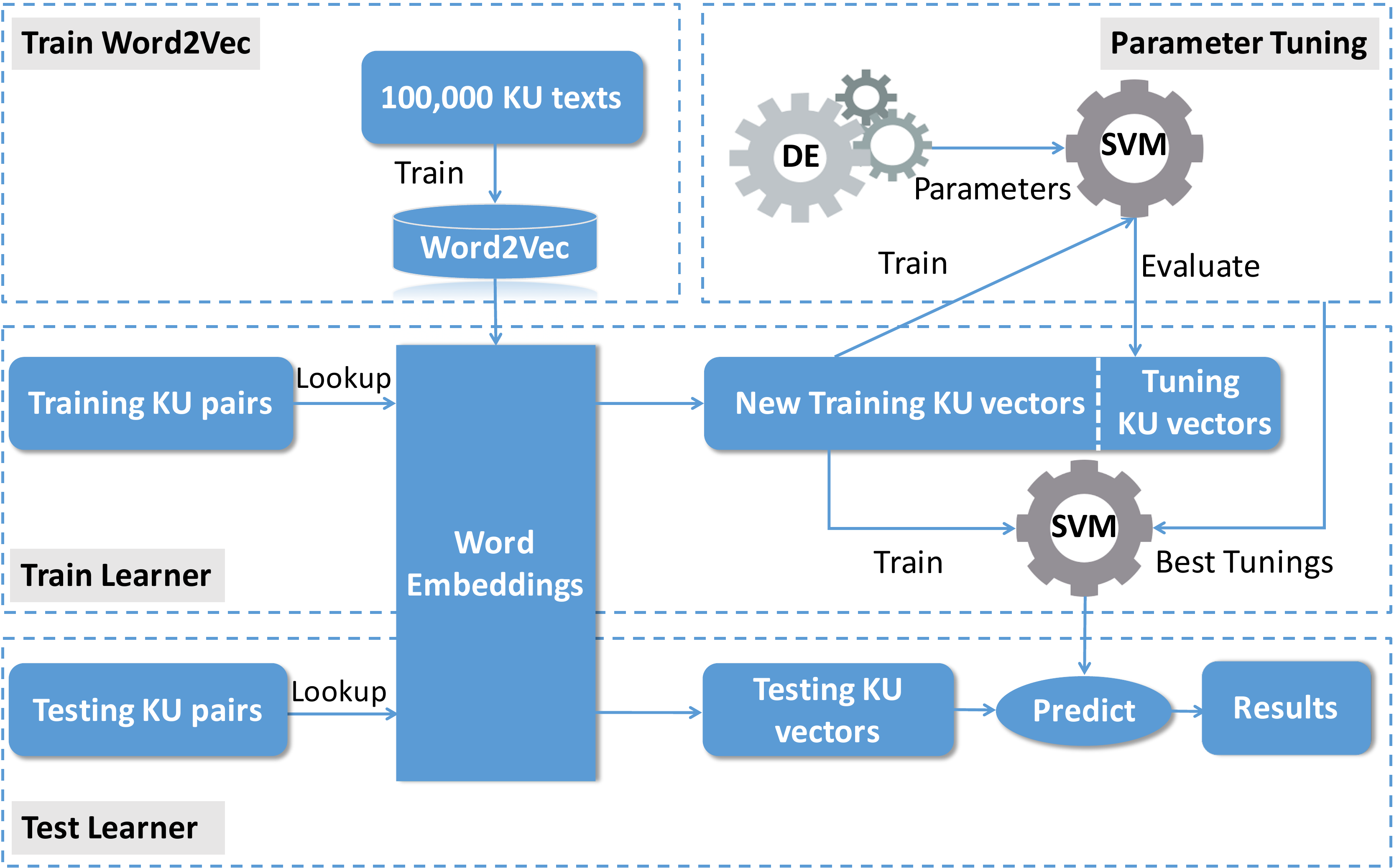} 
    \caption{The Overall Workflow of Building Knowledge Units Predictor with Tuned SVM}
    \label{fig:workflow}
\end{figure}

As seen in \fig{workflow}, instead of using all the 6,400 knowledge units as training data, 
we split the original training data into { new training data} and {tuning data}, which are
used during parameter tuning procedure for training SVM and evaluating candidate
parameters offered by DE. Afterwards, the {new training} data is again fitted into the SVM
with the optimal parameters found by DE and finally  the performance of the tuned
SVM will be evaluated on the {testing data}.

To reduce the potential variance caused
by how the original training data is divided, {\it 10-fold cross-validation} is performed. Specifically, 
each time one fold with $640$ knowledge units pairs is used as the tuning data, and the remaining folds with $5760$
knowledge units are used as  the new training data, then the output SVM model will be evaluated on the testing data. Therefore,
all the performance scores reported below are averaged values over 10 runs.

In this study, we use Wilcoxon single ranked test to statistically compare
the differences between tuned SVM and untuned SVM.
Specifically, the Benjamini-Hochberg~(BH) adjusted p-value 
is used to test whether a difference is statistically significant
at the level of $0.05$~\cite{benjamini1995controlling}. 
To measure the effect size of performance scores between tuned SVM and untuned SVM,
we compute Cliff's $\delta$ that is a non-parametric effect size measure~\cite{romano2006exploring}.
As Romano et al. suggested, we evaluate the magnitude of the effect size as follows:
negligible ($|\delta|<0.147$ ), small ($ 0.147<|\delta|<0.33$), medium ($0.33<|\delta|<0.474$ ), and large (0.474 $\leq|\delta|$)~\cite{romano2006exploring}.

\subsection{Evaluation Metrics}

When evaluating the performance of tuning SVM on the
multi-class linkable knowledge units prediction problem,
consistent with XU~\cite{xu2016predicting}, we use accuracy, precision, recall and F1-score
as the evaluation metrics.

\begin{table}[htp]
\caption {Confusion Matrix.}
\scriptsize
\resizebox{0.32\textwidth}{!}{
\begin{tabular} {@{}cc|c|c|c|c|l@{}}
\cline{3-6}
& & \multicolumn{4}{ c| }{Classified as} \\ \cline{3-6}
& & $C_1$ & $C_2$ & $C_3$ & $C_4$ \\ \cline{1-6}
\multicolumn{1}{ |c  }{\multirow{4}{*}{\rotatebox[origin=c]{90}{Actual}} } &
\multicolumn{1}{ |c| }{$C_1$} & $c_{11}$ & $c_{12}$  & $c_{13}$ & $c_{14}$  & \\ \cline{2-6}
\multicolumn{1}{ |c  }{}                        &
\multicolumn{1}{ |c| }{$C_2$} & $c_{21}$& $c_{22}$ & $c_{23}$ & $c_{24}$ &  \\ \cline{2-6}
\multicolumn{1}{ |c  }{}                        &
\multicolumn{1}{ |c| }{$C_3$} & $c_{31}$ & $c_{32}$ & $c_{33}$ & $c_{34}$ & \\ \cline{2-6}
\multicolumn{1}{ |c  }{}                        &
\multicolumn{1}{ |c| }{$C_4$} & $c_{41}$ & $c_{42}$ & $c_{43}$ & $c_{44}$ & \\ \cline{1-6}
\end{tabular}}

\label{tab:confusion}
\end{table}

Given a multi-classification problem with true labels $C_1$, 
$C_2$, $C_3$ and $C_4$, we can generate a confusion matrix like \tab{confusion}, 
where the value of $c_{ii}$ represents the number of instances that are correctly classified
by the learner for class $C_i$. 

{\it Accuracy} of the learner is defined as the number of  correctly
classified knowledge units over the total number of knowledge units, i.e.,

{\[
\begin{array}{ll}
accuracy = \frac{\sum_i c_{ii}}{\sum_{i}\sum_{j}c_{ij}}
\end{array}
\]}
where ${\sum_{i}\sum_{j}c_{ij}}$ is the total number of knowledge units.
For a given type of knowledge units, $C_j$, the  precision is defined as probability of
knowledge units pairs correctly classified as $C_j$ over the number of knowledge unit pairs classified as $C_j$ and
 recall is defined as the percentage of all $C_j$ knowledge unit pairs correctly classified. F1-score is the harmonic mean of
 recall and precision. Mathematically,
  precision, recall and  F1-score of 
the learner for class $C_j$ can be denoted as follows:

{\[
\begin{array}{ll}
prec_j &= precision_j = \frac{c_{jj}}{\sum_{i}c_{ij}}\\
pd_j &= recall_j = \frac{c_{jj}}{\sum_{i}c_{ji}}\\ 
F1_{j} &= 2*pd_j*prec_j/(pd_j + prec_j)
\end{array}
\]}
Where $\sum_{i}c_{ij}$ is the predicted number of knowledge units in class $C_j$ 
and ${\sum_{i}c_{ji}}$ is the actual number of knowledge units in class $C_j$.

Recall from Algorithm~1 that we call differential evolution once for each
optimization goal. Generally, this goal depends on which metric is most important for
the business case. In this work, we use $F1$ to score the candidate parameters because it controls
the trade-off between precision and recall, which is also consistent with XU~\cite{xu2016predicting}
and is also widely used in software engineering
community to evaluate classification results~\cite{wang2016automatically,menzies2007data,fu2016tuning,kim2008classifying}.

\section{Results}\label{results}
In this section, we present our experimental results. To answer research questions raised in
Section~\ref{RQ}, we conducted two experiments:

\begin{figure}[!htp]
    \centering
     \includegraphics[width=0.49\textwidth,height=2.2in]{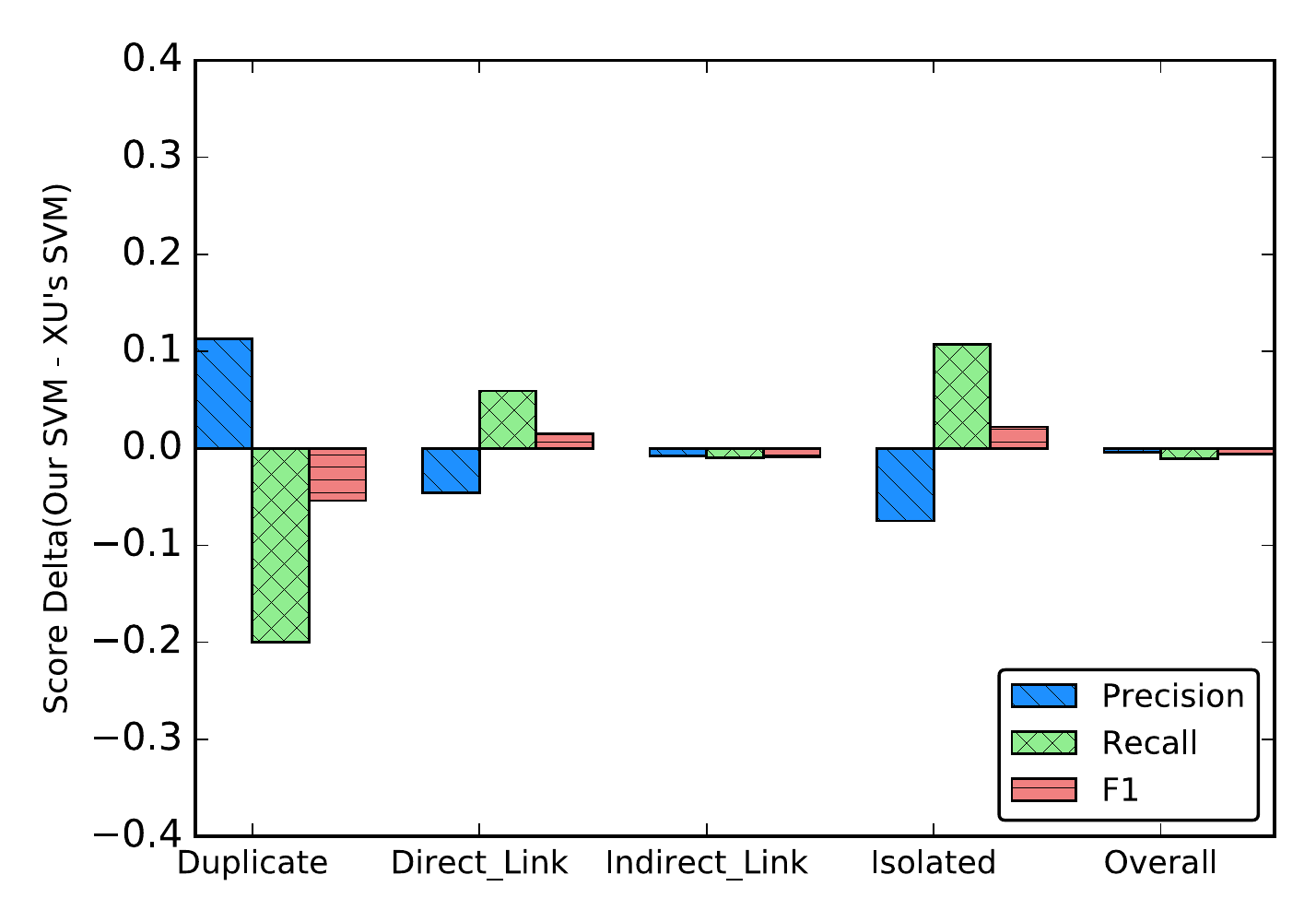} 
    \caption{Score Delta between Our SVM with XU's SVM in ~\cite{xu2016predicting} in Terms of Precision, Recall and F1-score. Positive Values Mean
            Our SVM is Better than XU's SVM in Terms of Different Measures; Otherwise, XU's SVM is better.}
    \label{fig:OurSVM-Xu'sSVM}
\end{figure}

\bi
\item
Compare  performance of Word Embedding + SVM method in XU~\cite{xu2016predicting} and our implementation;
\item
Compare performance  of our tuning SVM with DE method with XU's CNN deep learning method.

\ei
Since we used the same training and testing data sets provided by XU~\cite{xu2016predicting} and conducted our experiment in the same procedure and evaluated methods using the performance measures, we simply used the results reported in the work by XU~\cite{xu2016predicting} for the performance comparison.


\textbf{RQ1: Can we reproduce XU's baseline results (Word Embedding + SVM)?}

This first question   is    important to our work since, without the original tool released by XU,
we need to insure that  our reimplementation of their baseline method~(WordEmbedding + SVM) has a similar performance to their work.
Accordingly, we carefully follow XU's procedure~\cite{xu2016predicting}. We 
use the SVM learner from scikit-learn with the setting $\gamma = \frac{1}{200}$ and $\mathit{kernel=}$``rbf'', which are used by XU. After that, the same training and testing knowledge unit pairs are applied to SVM.

\begin{table}[!htp]
\centering
\caption{Comparison of Our Baseline Method with XU's. The Best Scores are Marked in \textbf{Bold}.}
\resizebox{0.48\textwidth}{!}{
\begin{tabular} {@{}l l  c c  c c c@{}}
\hline
   \multirow{2}{*}{Metrics} &  \multirow{2}{*}{Methods} &  \multirow{2}{*}{Duplicate} &  
   \multirow{2}{*}{\begin{tabular}[c]{@{}c@{}}Direct \\ Link\end{tabular}} &
   \multirow{2}{*}{\begin{tabular}[c]{@{}c@{}}Indirect \\ Link\end{tabular}} & 
   \multirow{2}{*}{Isolated} &  \multirow{2}{*}{Overall} \\ \\ \hline
   \multirow{2}{*}{Precision}
    & Our SVM &\textbf{0.724} &0.514 & 0.779 &0.601& 0.655 \\
   & XU's SVM &0.611 &\textbf{0.560} &\textbf{0.787}&\textbf{0.676}&\textbf{0.659} \\ \hline
   \multirow{2}{*}{Recall} 
   & Our SVM & 0.525& \textbf{0.492} & 0.970 & \textbf{0.645}  & 0.658 \\ 
   & XU's SVM  & \textbf{0.725} &0.433    &\textbf{0.980}  & 0.538 &\textbf{0.669}  \\ \hline
   \multirow{2}{*}{F1-score}
   & Our SVM & 0.609 &  \textbf{0.503} &0.864  &  \textbf{0.622}&0.650 \\ 
   & XU's SVM & \textbf{0.663} &  0.488  & \textbf{0.873} &  0.600 &\textbf{0.656} \\ \hline
   \multirow{2}{*}{Accuracy} 
   &  Our SVM&0.525  &  0.493 & 0.970 & 0.645 &0.658 \\
   & XU's SVM & - &  -  &- &  - &\textbf{0.669} \\ \hline
 \end{tabular}}
\label{tab:baseline}
\end{table}

 \begin{figure}[!htp]
    \centering
    \includegraphics[width=0.49\textwidth,height=2.2in]{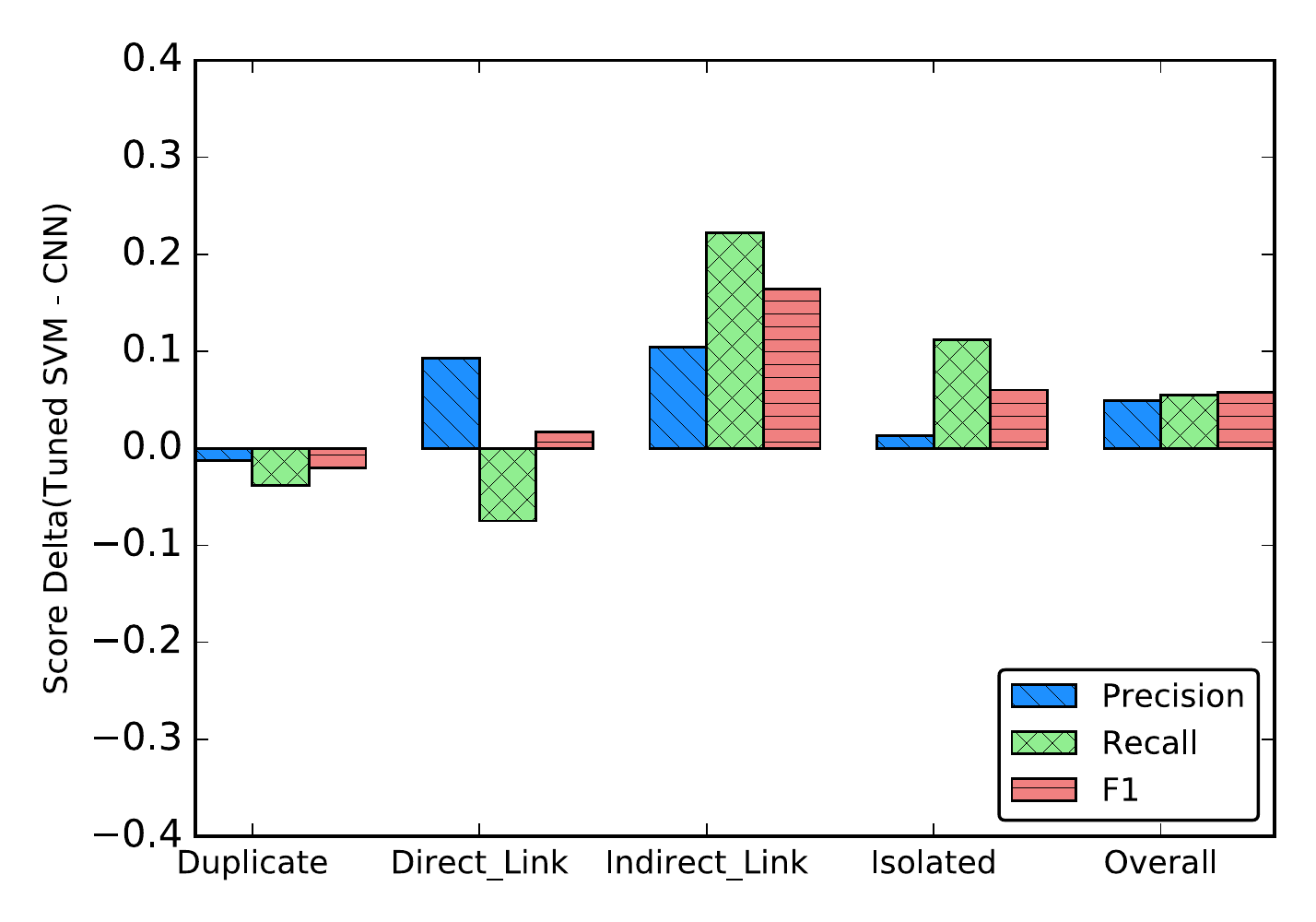} 
    \caption{Score Delta between Tuned SVM and CNN method~\cite{xu2016predicting} in Terms of Precision, Recall and F1-score. Positive Values Mean
            Tuned SVM is Better than CNN in Terms of Different Measures; Otherwise, CNN is better.}
    \label{fig:TunedSVM-CNN}
\end{figure}

 \tab{baseline}  and \fig{OurSVM-Xu'sSVM} show the performance scores and corresponding score delta between our implementation of WordEmbedding + SVM  with
 XU's in terms of accuracy~\footnote{XU just report overall accuracy, not for each class, hence it is missing in this table.}, precision, recall and F1-score. As we can see, 
 when predicting these four different types of relatedness between knowledge unit pairs,
 our Word Embedding + SVM method has  very  similar performance scores to the baseline method
 reported by XU in ~\cite{xu2016predicting}, with the maximum difference less than $0.2$.  
 Except for { Duplicate} class, where our baseline 
has a higher precision~(i.e., $0.724$ v.s. $0.611$) but a lower  recall (i.e., $0.525$ v.s.$0.725$).

\fig{OurSVM-Xu'sSVM} presents the same results in a graphical format.
Any bar above zero means that our implementation has a better performance score than XU's on predicting that specific knowledge unit relatedness class. As we can see, most 
of the differences ($\frac{8}{12}$) are within 0.05 and the score delta of overall performance shows that our
implementation is a little worse than XU's implementation.
For this chart we conclude that:

\vskip 1ex
 \begin{myshadowbox}
Overall, our reimplementation of WordEmbedding + SVM
has very similar performance in all the evaluated metrics 
 compared to the baseline method reported in XU's study ~\cite{xu2016predicting}.

 \end{myshadowbox}
The significance of this conclusion is that, moving forward,
we are confident that we can  use our  reimplementation of WordEmbedding+SVM as  a valid surrogate for
the baseline method of XU.

\textbf{RQ2: Can DE tune a standard learner such that it outperforms XU's deep learning method?}

To answer this question, we run the workflow of \fig{workflow}, where DE is applied to 
find the optimal parameters of SVM based on the training and tuning data. The optimal tunings are then applied on the SVM model and  the built learner is evaluated on testing data. Note that, in this study,
since we mainly focus on {precision}, {recall} and {F1-score} measures where {F1-score} is the
harmonic mean of {precision} and {recall}, we use {F1-score} as the tuning goal for DE.
In other words, when tuning parameters,  DE expects to find a pair of candidate parameters that maximize
 F1-score. 

 \begin{figure}[!t]
    \centering
     \includegraphics[width=0.49\textwidth,height=2.2in]{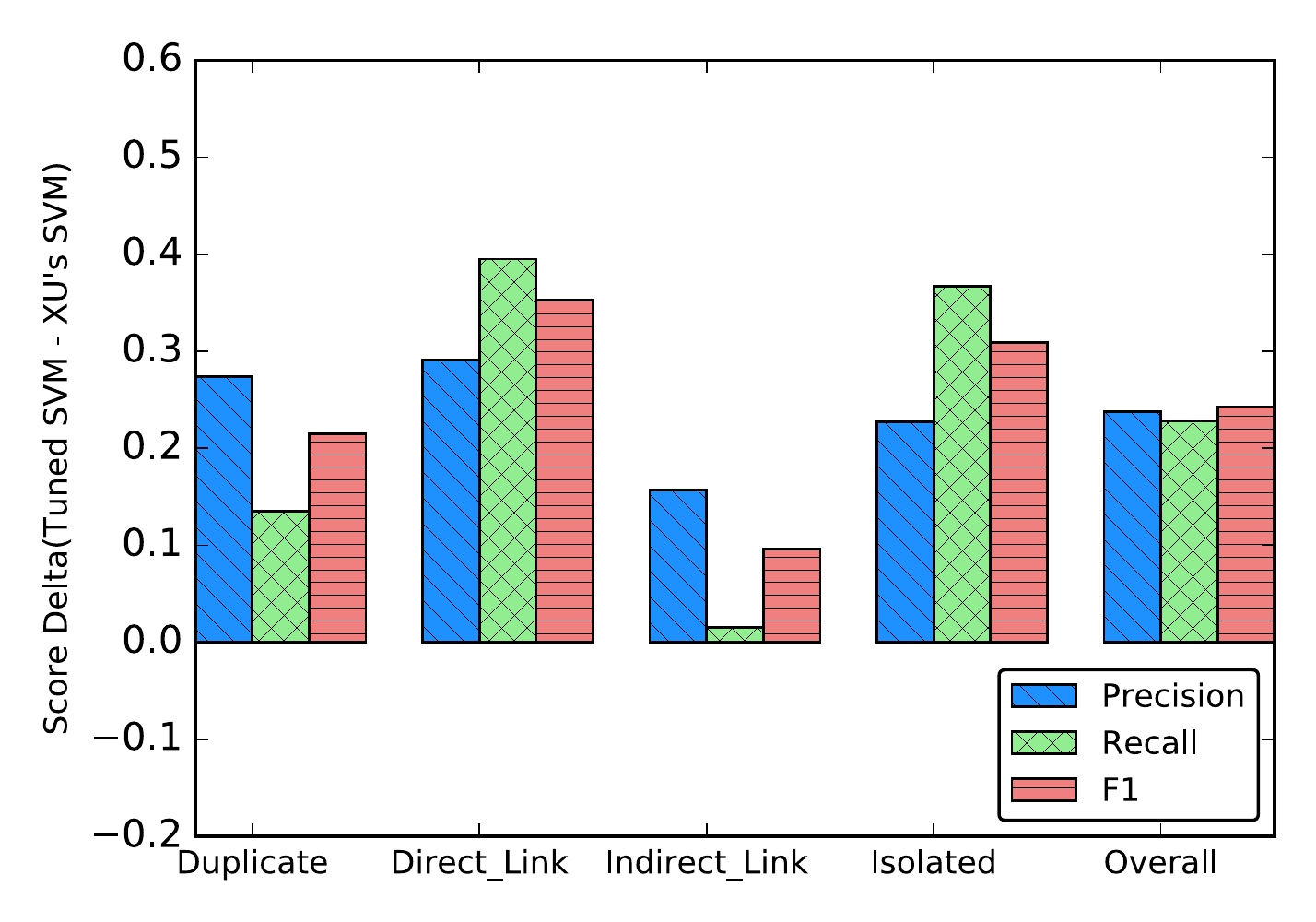} 
    \caption{Score Delta between Tuned SVM and XU's Baseline SVM in Terms of Precision, Recall and F1-score. Positive Values Mean
            Tuned SVM is Better than XU's SVM in Terms of Different Measures; Otherwise, XU's SVM is better.}
    \label{fig:TunedSVM-SVM}
\end{figure}

\begin{table}[!htp]
\centering
\caption{Comparison of Tuned SVM with XU's CNN Method. The Best Scores are Marked in \textbf{Bold}. }
\resizebox{0.48\textwidth}{!}{
\begin{tabular} {@{}l l  c c  c c c@{}}
\hline
   \multirow{2}{*}{Metrics} &  \multirow{2}{*}{Methods} &  \multirow{2}{*}{Duplicate} &  
   \multirow{2}{*}{\begin{tabular}[c]{@{}c@{}}Direct \\ Link\end{tabular}} &
   \multirow{2}{*}{\begin{tabular}[c]{@{}c@{}}Indirect \\ Link\end{tabular}} & 
   \multirow{2}{*}{Isolated} &  \multirow{2}{*}{Overall} \\ \\ \hline
  \multirow{3}{*}{Precision} 
   & XU's SVM&0.611 &0.560 &0.787&0.676&0.658 \\ 
   & XU's CNN&\textbf{0.898} & 0.758&0.840 &0.890 &0.847 \\
   & Tuned SVM&0.885 & \textbf{0.851}&\textbf{0.944} &\textbf{0.903} &\textbf{0.896}\\ \hline
   \multirow{3}{*}{Recall} 
   & XU's SVM& 0.725 &0.433    &0.980  & 0.538 &0.669  \\
   & XU's CNN& \textbf{0.898}&\textbf{0.903}    &0.773  & 0.793 &0.842  \\ 
   & Tuned SVM& 0.860 &0.828    &\textbf{0.995}  & \textbf{0.905} &\textbf{0.897}  \\  \hline
   \multirow{3}{*}{F1-score}
   & XU's SVM& 0.663 &  0.488  & 0.873 &  0.600 &0.656 \\ 
   & XU's CNN& \textbf{0.898} &  0.824  & 0.805 &  0.849 &0.841 \\
   & Tuned SVM& 0.878 & \textbf{ 0.841}  &\textbf{ 0.969} &  \textbf{0.909} &\textbf{0.899} \\\hline
 \end{tabular}}
\label{tab:RQ2}
\end{table}

\tab{RQ2} presents the performance scores of  XU's baseline, XU's CNN method and Tuned SVM 
for all metrics. The highest score for each relatedness class
  is marked in bold.  Note that: Without tuning, XU's CNN method outperforms
the baseline SVM in $\frac{10}{12}$ evaluation metrics across all four classes. 
The largest performance improvement is $0.47$ for {recall} on {Direct Link} class. Note that this result is consistent with XU's conclusion
that their CNN method is superior to standard SVM.
After tuning SVM, the deep learning method has no such advantage. Specifically, CNN has advantage over tuned SVM in $\frac{4}{12}$ evaluation metrics across all four classes. Even when CNN performs better that our tuning SVM method,
the largest difference is $0.065$ for Recall on Direct\_Link class, which is less than $0.1$. 

\fig{TunedSVM-CNN} presents the same results in a graphical format.
Any bar above zero indicates that tuned SVM has
a better performance score than CNN.
In this figure: CNN has a slightly better performance on
{Duplicate} class for {precision}, {recall} and {F1-score} and
a higher {recall} on {Direct link} class.  Across all of \fig{TunedSVM-CNN},
in $\frac{8}{12}$ evaluation scores, Tuned SVM has better performance scores than CNN, with the largest delta of $0.222$.

\begin{figure}[!t]
    \centering
     \includegraphics[width=0.49\textwidth,height=2.1in]{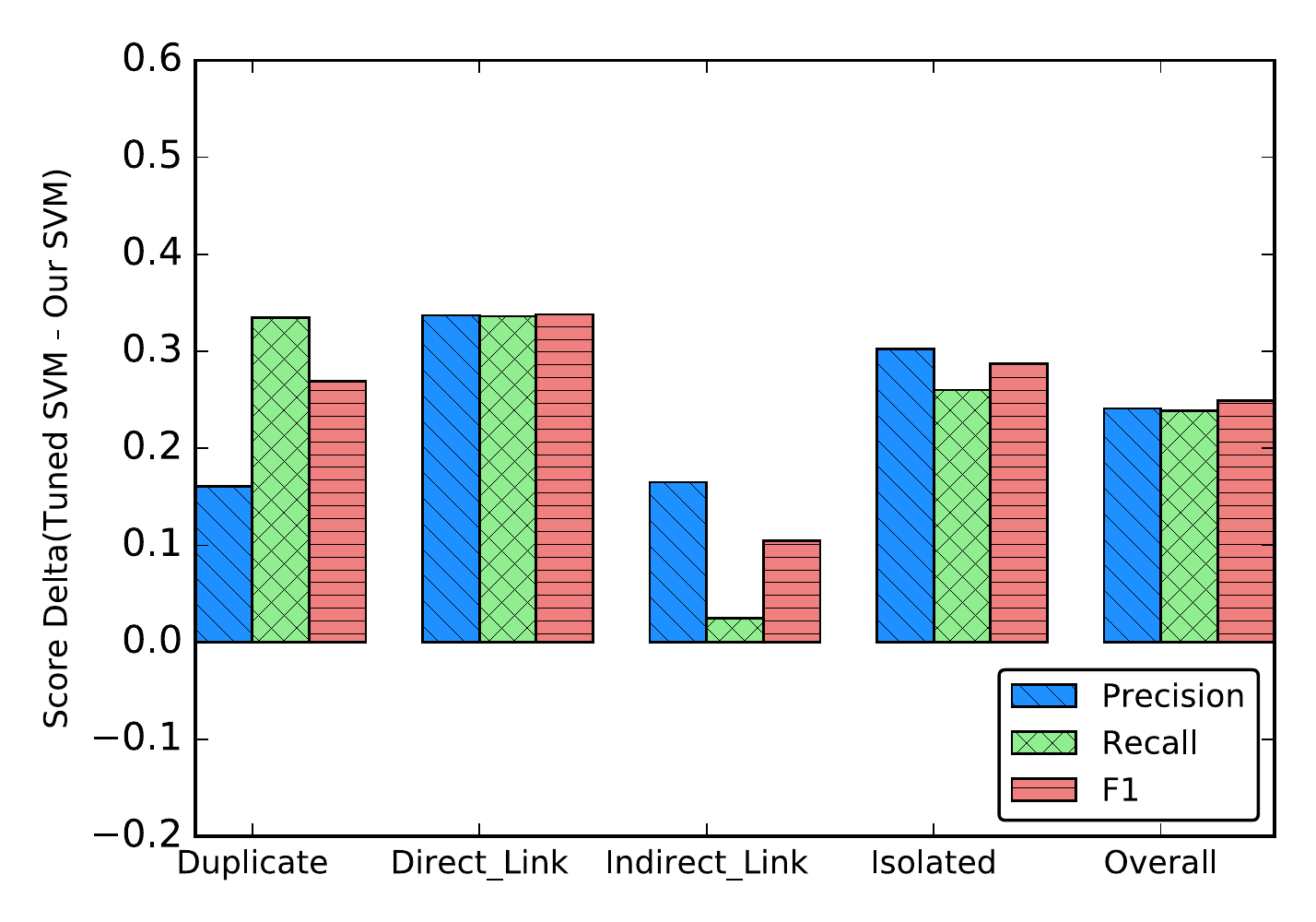} 
    \caption{Score Delta between Tuned SVM and Our Untuned SVM in Terms of Precision, Recall and F1-score. Positive Values Mean
            Tuned SVM is Better than Our Untuned SVM in Terms of Different Measures; Otherwise,  Our  SVM is Better.}
    \label{fig:TunedSVM-OurSVM}
\end{figure}

\fig{TunedSVM-SVM} compares the performance delta of tuned SVM with XU's untuned SVM.
 We note that DE-based parameter tuning never degrades SVM's performance
 (since there are no negative values in that chart). 
 Tuning dramatically improves scores on predicting some classes of KU relatedness. 
 For example, the {recall} of predicting {Direct\_link} is increased 
 from $0.433$ to $0.903$, which is $108\%$ improvement over XU's untuned SVM
 (To be fair for XU, it is still $84\%$ improvement over our untuned SVM).
 At the same time, the corresponding {precision} and {F1} scores of predicting {Direct\_Link} 
 are increased from $0.560$ to $0.851$ and $0.488$ to $0.841$, 
which are $52\%$ and $72\%$ improvement over XU's original report\cite{xu2016predicting}, respectively. 
 A similar pattern can also be observed in {Isolated} class. On average, tuning helps improve the performance
of XU's SVM by $0.238$, $0.228$ and $0.227$ in terms of {precision}, {recall} and {F1-score}
for all four KU relatedness classes. \fig{TunedSVM-OurSVM} compares the tuned SVM with our untuned SVM. We note that we get 
the similar patterns that observed in \fig{TunedSVM-SVM}. All the bars are above zero, etc.


Based on the performance scores in \tab{RQ2} and score delta in \fig{TunedSVM-CNN}, \fig{TunedSVM-SVM} and \fig{TunedSVM-OurSVM},
 we can see that:
 \bi
 \item
 Parameter tuning can dramatically improve the performance of Word Embedding + SVM~(the baseline method) for the multi-class KU relatedness prediction task;
\item
With the optimal tunings, the traditional machine learning method, SVM, if not better, is at least comparable 
 with deep learning methods (CNN).
 \ei
When discussing this result with colleagues, we are sometimes asked for a statistical
analysis that confirms the above finding.
However, 
 due the lack of evaluation score distributions of the CNN method in~\cite{xu2016predicting}, we cannot compare
their single value with our results from 10 repeated runs. However, according to Wilcoxon singed rank test over 10 runs results, tuned SVM performs
statistically better than our untuned SVM in terms of all evaluation measures on all four classes~($p<0.05$).
According to Cliff $\delta$ values, the 
magnitude of difference between tuned SVM and our untuned SVM is not trivial~($|\delta| > 0.147$) for all evaluation
measures. 
 
 Overall, the experimental results and our analysis indicate that:

\vskip 1ex
 \begin{myshadowbox}
In the evaluation conducted here,
the deep learning method, CNN, does not have any performance advantage over our tuning approach.
 \end{myshadowbox}

\textbf{ RQ3: Is tuning SVM with DE faster than XU's deep learning method?}

When comparing the runtime of two learning methods, it obviously should be conducted under the
 same hardware settings. 
 Since we adopt the CNN evaluation scores from ~\cite{xu2016predicting},
 we can not run on our tuning SVM experiment under the exactly same system settings.
 To allow readers to have a objective comparison, we provide the experimental environment as shown in~\tab{env}. 
 To obtain the runtime of tuning SVM, we  recorded the start time and end time
 of the program execution, including parameter tuning, training model and testing model.

 \begin{table}[!h]
\centering
\caption{Comparison of Experimental Environment  }
\begin{tabular} {@{}l  l l l @{}}
\hline
Methods&OS&CPU&RAM \\ \hline
Tuning SVM & MacOS 10.12 & Intel Core i5 2.7 GHz & 8 GB  \\
CNN& Windows 7 &Intel Core i7 2.5 GHz & 16 GB \\
\hline
\end{tabular}
\label{tab:env}
\end{table}

According to XU, it took  $14$ hours to train their  CNN model
into a low loss convergence~($< e^{-3}$)~\cite{xu2016predicting}.
Our work, on the other hand
only takes $10$ minutes to run SVM with parameter tuning
by DE on a similar environment.
That is, the simple parameter tuning method on SVM is $84X$ faster than  XU's deep learning method.
 
\vskip 1ex
 \begin{myshadowbox}
Compared to CNN method, tuning SVM is about $84X$ faster in terms of model building.

 \end{myshadowbox}
The significance of this finding is that, in this case study,
CNN was neither better in performance scores (see RQ2) nor runtimes.
CNN's extra runtimes are a particular concern since (a) they are very long;
and (b)~these would be incurred anytime researchers wants to update the CNN model
with new data or wanted to validate the XU result.

\section{Discussion}\label{discussion}

\subsection{Why DE+SVM works?}

\textbf{Parameter tuning matters}. As mentioned in Section \ref{tuning}, the default parameter values set by the algorithm designers could generate a good performance on average but may not guarantee the best performance for the local data~\cite{bergstra2012random,fu2016tuning}. Given that,
it is most strange to report that  most SE researchers ignore the impacts of
parameter tuning when they utilize various machine learning methods to conduct software analytic (evidence: see our reviews in \cite{fu2016tuning,fu2016differential,agrawal2016wrong}). 
The conclusion of this work must be to stress the importance of this kind of tuning, using local data, for any future software analytics study.

\textbf{Better explore the searching space}.  It turns out that one exception to our statement that ``most researchers do not tune''
is the XU study. In that work, they unsuccessfully perform parameter tuning, but with with grid search. In such a grid search, for $N$ parameters to be tuned, $N$ for loops are created
to run over a range of settings for each parameter. While a widely used method,  it is often deprecated. For example,  Bergstra et al.\cite{bergstra2012random} note that  grid search jumps through different parameter settings between some {\it min} and {\it max} values of pre-defined tuning range. They warn that such jumps may actually
skip over the  critical tuning values. On the other hand, DE tuning values are adjusted based on better candidates from previous generations. Hence DE is more likely than grid search to ``fill in the gaps'' between the initialized values. 

That said, although DE +SVM works in this study, it does not mean DE is the best parameter tuner for all SE tasks. We encourage more researchers to explore faster and more effective parameter tuners in this direction.
 
\subsection{Implication}

Beyond the specifics of this case study, what general principles
can we take from the above work?

\textbf{Understand the task.}
One reason to try different tools for the same task is to better
understand the task.
The more we understand a task, the better we can match tools to that task. Tools that are poorly matched to task are usually complex and/or slow to execute.  In the case study of this paper, we would say that

\bi
\item
Deep learning is a poor match to the task of predicting whether two questions posted on Stack Overflow are semantically linkable
since it is so slow;
\item
Differential evolution tuning SVM is a much better match since it is so fast and obtain competitive performance.
\ei
That said, 
it  is important to stress that the point of this study
is not to deprecate deep learning.  
There are many scenarios were we
 believe  deep learning would be a natural
 choice (e.g. when analyzing complex speech or visual data). 
 In SE, it is still an open research question that in which scenario deep learning
 is the best choice. Results from this paper show that, at least for classification tasks like 
 knowledge unit relatedness classification on Stack Overflow,
 deep learning does not have much advantage over well tuned conventional machine learning methods. 
However, as we better understand SE tasks, deep learning could be used to address more SE problems,
which require more advanced artificial intelligence.

\textbf{Treat resource constraints as design challenges.}
As a general engineering principle,
we think it insightful to consider the resource cost
of a tool before applying it.
It turns out that this is a design pattern used in contemporary industry.
According to Calero and Pattini~\cite{calero2015green},  many current commercial  redesigns are motivated (at least in part) by arguments based on sustainability (i.e. using fewer resources to achieve results).
In fact, they say that managers used sustainability-based redesigns to 
motivate extensive cost-cutting opportunities.

\subsection{Threads to Validity}

Threats to \textbf{internal validity} concern the consistency of the results 
obtained from the result. In our study,  to investigate how
tuning can improve the performance of baseline methods and how well
it perform compared with deep learning method. We select
XU's  Word Embedding + SVM baseline method as a case study. Since the original implementation of 
Word Embedding + SVM (baseline 2 method in ~\cite{xu2016predicting}) is not 
publicly available, we have to reimplement our version of Word Embedding + SVM as
the baseline method in this study. As shown in RQ1, our implementation has
quite similar results to XU's on the same data sets. Hence, we believe that our implementation reflect the original
 baseline method in Xu's study~\cite{xu2016predicting}. 
 

 Threats to \textbf{external validity} represent if the results are of relevance for
 other cases, or the ability to generalize the observations in a study. In this study,
 we compare our tuning baseline method with deep learning method, CNN, in terms of
 precision, recall, F1-score and accuracy. The experimental results are quite consistent
 for this knowledge units relatedness prediction task. 
 Nonetheless, we do not claim that our findings can be generalized to all software analytics tasks. 
 However, those other software analytics tasks often apply deep learning
 methods on classification tasks~\cite{choetkiertikul2016deep, wang2016automatically} 
 and so it is quite possible that
 the methods  of this paper (i.e., DE-based parameter tuning) would
 be widely applicable, elsewhere.

\section{Conclusion}\label{conclusion}

In this paper, we perform a comparative study to investigate
how tuning can improve the baseline method compared with
 state-of-the-art deep learning method  for predicting
knowledge units relatedness on Stack Overflow. Our experimental
results show that:

\bi
\item Tuning improves the performance of baseline methods. 
At least for Word Embedding + SVM~(baseline in ~\cite{xu2016predicting}) method, if not better,
it performs as well as the proposed CNN method in ~\cite{xu2016predicting}.
\item The baseline method with parameter tuning runs much faster than complicated deep learning.
In this study, tuning SVM runs $84X$ faster than CNN method.
\ei

\section{Addendum}
As this paper was going to going to press we learned of a new deep learning methods that, according to its
creators, runs 20 times faster than standard deep learning~\cite{spring2016scalable}. 
Note that in that paper, the authors say  their faster method does not produce better results-- 
in fact, their method generated solutions that were a small fraction worse than ``classic'' deep learning. Hence, that paper does not invalidate our result since (a)~our DE-based
method sometimes produced {\em better} results than classic deep learning and (b)~our DE runs 84 times faster (i.e. much faster runtimes than those reported in~\cite{spring2016scalable}).

That said, this new fast deep learner deserves our close attention since, using it, we conjecture that our DE tools
could solve an open problem in the deep learning community; i.e. how to find the best configurations inside a deep learner faster.


Based on the results of this study, we recommend that before applying 
deep learning method on SE tasks, implement
  simpler techniques.
  These simpler methods could be used,
  at the very least, for comparisons against a baseline.
  In this particular case of deep learning vs DE, the extra computational effort  is so very minor (10 minutes on top of 14 hours), 
  that such a ``try-with-simpler'' should
  be standard practice.
  

As to the future work, we will explore more simple techniques to solve SE tasks and also
investigate how deep learning techniques could be applied effectively in software engineering
field. 

\section*{ACKNOWLEDGEMENTS}
The work is partially funded by an NSF award \#1302169.

\balance
\bibliographystyle{plain}
\bibliography{easierthanDL.bbl}

\end{document}